\documentclass[%
 reprint,onecolumn,
nofootinbib,
 amsmath,amssymb,
 aps,
]{revtex4-2}

\usepackage{graphicx}% Include figure files
\usepackage{dcolumn}% Align table columns on decimal point
\usepackage{bm}% bold math
\usepackage{xcolor}
\usepackage{hyperref}
\usepackage{cancel}

\begin{document}

\newtheorem{theorem}{Theorem}
\newtheorem{definition}{Definition}
\newtheorem{lemma}{Lemma}
\newtheorem{proposition}{Proposition}
\newtheorem{remark}{Remark}
\newtheorem{corollary}{Corollary}
\newtheorem{example}{Example}

%\preprint{APS/123-QED}

\title{Anomalous diffusion in porous fractal media}

% Force line breaks with \\

%\\
% please add all authors name and surname separated by comma
\author{Alexander Iomin\textsuperscript{1} and Trifce Sandev\textsuperscript{2,3,4}}
\address{\textsuperscript{1}\textit{Solid State Institute, Technion---Israel Institute of Technology, Haifa, Israel}}
\address{\textsuperscript{2}\textit{Research Center for Computer Science and Information Technologies, Macedonian Academy of Sciences and
Arts, Skopje, Macedonia}}
\address{\textsuperscript{3}\textit{Institute of Physics, Faculty of Natural Sciences and Mathematics, Ss. Cyril and Methodius University, Skopje, Macedonia}}
\address{\textsuperscript{4}\textit{Department of Physics, Korea University, Seoul, Republic of Korea}}

\date{\today}% It is always \today, today,
\begin{abstract}
We suggest a model of a diffusive process inside  a fractal sponge structure,
which is a generalization of the diffusion processes on a comb and fractal mesh structure.
The sponge model is considered as the direct product of Cantor sets.
It is shown that the  corresponding one-dimensional diffusion process is governed by a generalized Fokker-Planck equation with a power-law memory kernel and a position-dependent diffusion coefficient. That is, the fractal structure of the medium induces memory
effects and heterogeneity in the transport system. The considered model may be of interest to describe anomalous heat transport in porous fractal media.
\end{abstract}

\maketitle

\section{Introduction}\label{int}

\begin{figure}[h!]
(a) \includegraphics[width=8cm]{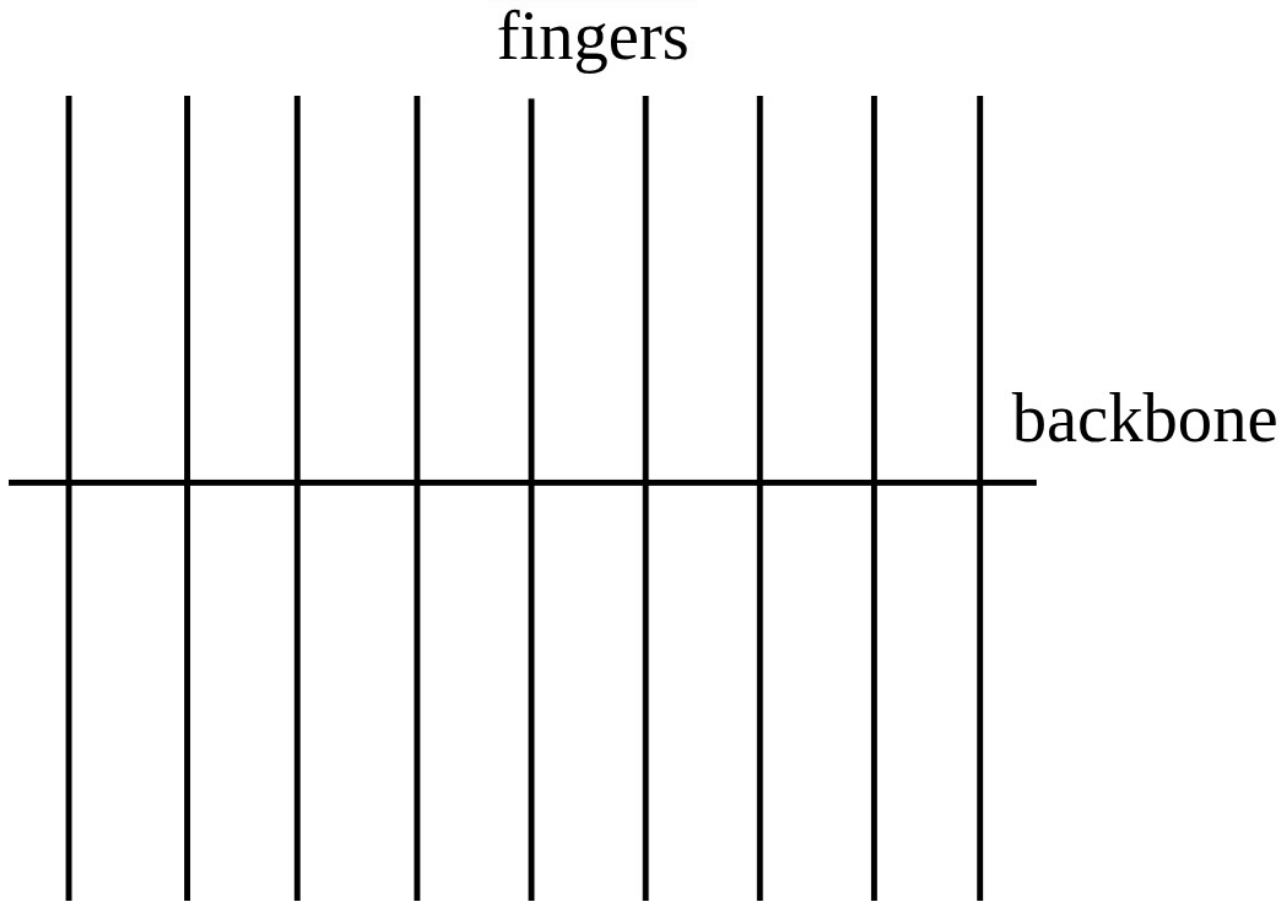} (b)\,\, \includegraphics[width=5.5cm]{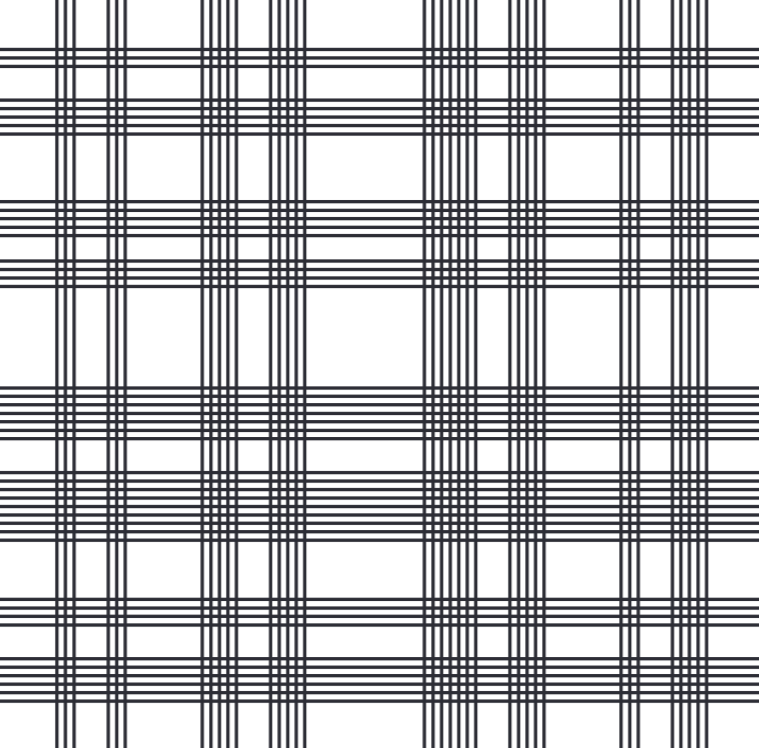}
\caption{ Fractal structure development from (a) comb to (b) fractal mesh 
(Cantor tartan).}\label{fig:comb}
\end{figure}

Transport in porous media is a long-lasting task, where the most efficient way of its description is a continuous (phenomenological) approach in 
porous media~\cite{bear72,BeBa90}. The implication of fractional calculus  results in significant progress in these studies of anomalous diffusion  in fractal porous media
that was eventually realized in fractional hydrodynamic equations of continuity~\cite{taras05a,taras05b} with further continuation of this study; see, \textit{e.g.},~\cite{ostoja2008,LiOs2009,CaSa2010,FoChHa2010,FoChHa2011a,FoChHa2011b,LiOs2013,LiOs2019,
Tar2020}. Among various tasks of this approach, a generalized 1D transfer equation has been introduced for ``Koch’s tree''-type fractal structure, as an example of fractal porous medium~\cite{nigmatulin86}, which reflects a memory effect due to fractal structure. It has the form of a fractional Fokker-Planck equation of order $1/2$. With variation 
of the geometry, this structure has been called a comb model with the amending statement  that such a comb-like structure can serve as a model of a porous medium~\cite{nigmatulin86}. Another completely independent consideration of the  comb model has been suggested in the field of percolation clusters~\cite{WhBa84,GeGo85,WeHa86}, 
where a random walker can move only onto conducting sites and is not 
allowed to step on non-conducting or isolating sites. At the percolation 
criticality, a percolation structure can be idealized as a single infinite 
cluster, consisting of a conducting path, which corresponds to a backbone, 
and side branches, also called teeth or fingers with dangling bonds.
In both cases, with some idealization, this structure corresponds to a 
comb shown in Figure~\ref{fig:comb}~(a).

An elegant mathematical realization of the comb geometry has been suggested 
in the form of the phenomenological Fokker-Planck equation~\cite{ArBa91}.
A matrix of diffusion coefficients reflects the comb geometry, which implies that displacement in the $x$-direction is possible only
along the structure axis, i.e., the $x$-axis at $y=0$.
In this way, diffusion in the $x$-direction is highly inhomogeneous, and
the diffusion coefficient determined by the Dirac $\delta$-function, which is $D_{xx}=D_x\delta(y)$, while the 
diffusion coefficient in the $y$-direction, the side-branch direction, is constant, that is $D_{yy}=D_y$.
Then the diffusion equation on the comb structure reads~\cite{ArBa91}
\begin{equation}\label{in-eq1}
\partial_tP(x,y,t)
=D_{x}\delta(y)\partial^{2}_xP(x,y,t)
+D_{y}\partial^{2}_yP(x,y,t),
\end{equation}
where $P(x,y,t)$ is the probability density function (PDF) 
of finding a diffusing particle at time $t$ at the
position with coordinates $(x,y)$ in the 2D comb space, while 
$\partial_r\equiv \frac{\partial}{\partial r}$, $r=\{x,y,t\}$. 
The comb model~\eqref{in-eq1} has been shown to be equivalent to the 
fractional Fokker-Planck equation of order $1/2$ for the marginal PDF $P_1(x,t)=\int_{-\infty}^{\infty}dy\, P(x,y,t)$, which describes subdiffusion along the backbone~\cite{IoBa05}.
Fractal generalizations of the comb model in the form of an inhomogeneous 
distribution of both the backbones and fingers have also been suggested, 
see Figure~\ref{fig:comb}~(b), and the corresponding anomalous transport has been studied~
\cite{iomin2011subdiffusion,sandev2015fractional,SaIoMe16,sandev2017anomalous,petreska2020time}.

%%%%%%%%%%%%%%%%%%%%%
%%%%%%%%%%

The comb models, introduced as a simplification approach to percolation 
clusters~\cite{WhBa84,GeGo85,WeHa86}, also capture the main signatures of anomalous transport 
in disordered systems. In this context, the comb model could be regarded as a geometric 
representation of the continuous time random walk~\cite{WeHa86,IoBa05,IoMeHo2018}. 
In this case, the fingers play the role of traps and are considered a 
synthetic dimension~\cite{synthDi1,synthD2}.
Various aspects of anomalous diffusion on combs, including a wide variety 
of possible applications, have a common feature of the classical random-walk process that in comb geometry leads to asymptotic subdiffusion along the backbone, with the mean-squared displacement (MSD) growing with time as $t^{1/2}$, see \textit{e.g.}, Refs.~\cite{IoMeHo2018,BoGe90,bAHa00,So12,SaIo2022,traytak2025,IoMiSa2026,liu2018comb,
liu2018anomalous}. This 
subdiffusive scaling law has been found in a variety of experimental situations and 
mathematical models.  In particular, 3D comb geometry with the corresponding 3D 
Fokker-Planck equation is used in microelectronics for the description of heat transport
in porous materials with low dielectric constant 
(low-$k$ material)~\cite{baklanov2003,baklanov2010,baklanov2025,baklanov2026}. 
In microelectronics, porous low–$k$ materials effectively reduce parasitic capacitance. 
However, their extremely low thermal conductivity creates severe heat-dissipation.  
This inefficient heat removal creates localized hot spots, accelerating electro-migration 
and dielectric degradation~\cite{baklanov2026,Cahill2003,Alam2014,Giri2020}.
Experimentally measured thermal conductivity shows a nonlinear decrease with increasing 
porosity~\cite{Alam2014}. This eventually means that thermal conductivity is a function of the 
fractal density of porous dielectric composites. According to a recent  
quantitative analysis of the fractal dimension performed in the framework of the 
Frenkel-Halsey-Hill model~\cite{PoTs2009}, the fractal dimension of the low-$k$ 
dielectric is $2.63$, see Ref.~\cite{baklanov2026}.
Therefore, 3D geometry is mandatory for the description of heat transport in porous low-$k$ dielectric materials. It should also be pointed out that along  with this heat transport kinetics, the electrodynamic properties of fractal composites 
have been extensively studied~\cite{StBeKo2004,SaSh2007,Tar05,MuBa2007,BaIo11a,BaIo11b,BaIo12,Nas2013,
BaIo13,BaMePaMo2013,Tar2015}, to name a few.

The main aim of the present research is to describe the heat transport characteristics 
of fractal artificial porous media, which can be considered as a possible 
theoretical counterpart of low-$k$ dielectric composites.
Without pretending to be a general theory of a complete explanation of 
heat transport in porous materials, we suggest a possible scenario of 
this phenomenon to estimate the transport characteristics as functions 
of the fractal dimension of porous fractal media.

The paper is organized as follows. In Section~\ref{rfs}, we consider a random fractal set as a product of three random Cantor sets, which will be used to model the structure of the fractal media (fractal sponge) in which the particles diffuse. The corresponding Fokker-Planck equation for a diffusing particle in a random fractal environment is introduced in Section~\ref{sde}. In Section~\ref{solut}, we present and discuss the analytical results for PDF and MSD. We show that anomalous diffusion is realized in the system due to the fractal structure of the sponge. Some limiting cases are also considered and analyzed to validate the obtained results. The summary is provided in Section~\ref{sum}. The definition, some properties, and asymptotic behavior of the Fox $H$-function are provided in an appendix at the end of the paper.

\section{Random fractal set}\label{rfs}

We consider a random fractal set as a direct product of random Cantor sets 
$S=S_{\alpha}\times S_{\beta}\times S_{\gamma}$, where each set has a fractal density
according to its fractal dimension $d$ as follows $\bar{\rho}(l)=\sum_{l_d\in S_d}\delta(l-l_d)$.
Here, $l=\{|x|,|y|,|z|\}$ coordinates  with the fractal dimensions $d=\{\alpha,\beta,\gamma\}$,
respectively. In the present analysis, we shall replace this singular fractal density with its continuous counterpart ~\cite{tar2004}, which relates to the integration of the PDF
 on the fractal volume\footnote{This replacement is justified and relates 
to the integration of a test function $p(l)$ on the fractal volume $\mu(l)$, that is, 
$\int p(l)\bar{\rho}(l)dl=\sum_{l_d\in S_d}p(l_d)\rightarrow 
\int p(l)d\mu(l)=\int\rho(l)p(l)dl$,
where $\rho(l)={l^{d-1}}/{\Gamma(d)}$. It should be admitted that in Sec.~\ref{sde},
the test function $p(l)$ is a well behaved probability function.
Note also that this continuity replacement is also related to the
approaches of Stillinger \cite{stillinger} and Wilson \cite{wilson}; see also  Refs. \cite{tar2025,chandel}.}.
That is,
\begin{equation}\label{density-1d}
    \bar{\rho}(l)=\sum_{l_d\in S_d}\delta(l-l_d)\rightarrow \rho(l)=\frac{1}{\Gamma(d)}l^{d-1}.
\end{equation}
Therefore, the density of the fractal set with the fractal dimension 
$d_f=\alpha+\beta+\gamma$ is 
\begin{equation}\label{density-3d}
    \rho_S(x,y,z)=\frac{|x|^{\alpha-1} |y|^{\beta-1}|z|^{\gamma-1}}
{\Gamma(\alpha)\Gamma(\beta)\Gamma(\gamma)}.
\end{equation}
Our aim is to find the transport characteristics of the fractal composite structure 
at the percolation in the $x$ direction. Therefore, we follow the 3D fractal comb strategy
as a generalization of our previous consideration\footnote{It should be noted that 
we use here only the density of the fractal structure Eq.~\eqref{density-3d} and neither the Weierstrass function~\cite{West90,sandev2015fractional} nor the Riesz integral~\cite{SaKiMa93,sandev2017anomalous}.} of the 2D 
fractal mesh~\cite{iomin2011subdiffusion,sandev2015fractional,sandev2017anomalous,IoMeHo2018}.

\begin{figure}[h!]
 \begin{center}
 \includegraphics[width=8cm]{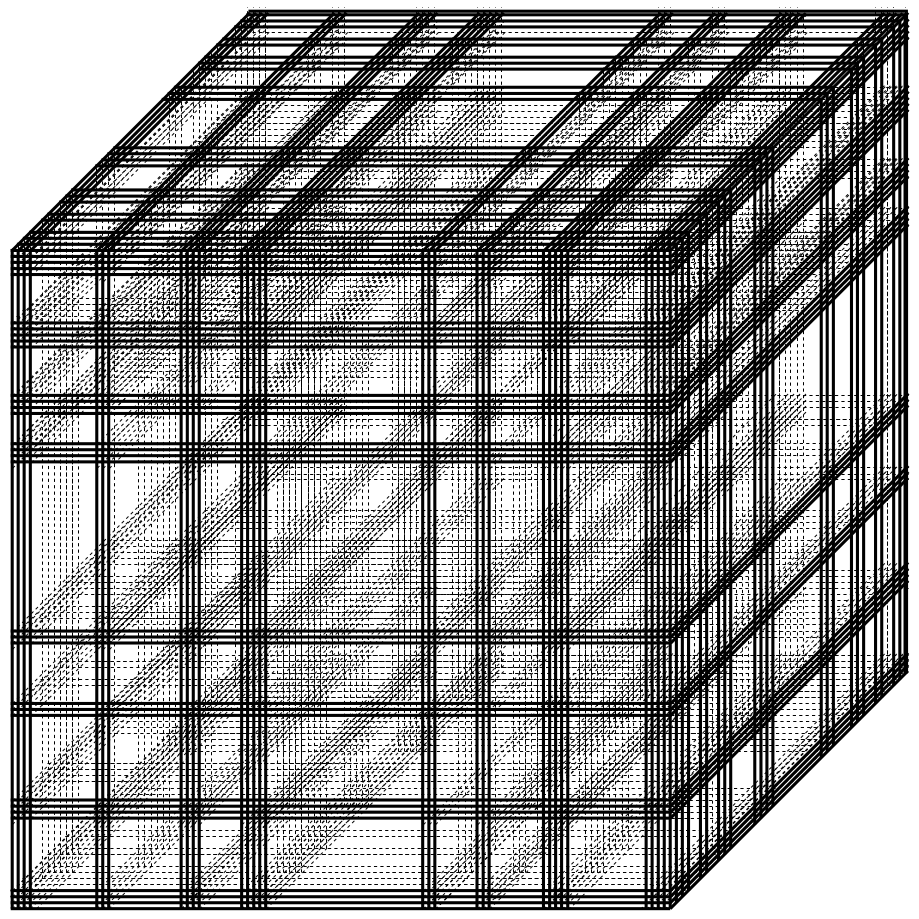}% This is a *.eps file
 \end{center}
 \caption{ A fractal sponge $S_{\alpha,\beta,\gamma}=S_{\alpha}\times S_{\beta}\times S_{\gamma}$, 
 which also defines a $3D$ fractal mesh of the diffusion coefficients matrix $D_{i,j}\neq 0$ in eq. \eqref{D-tensor}, if $(i,j)=\{x,y,z\}\in S_{\alpha,\beta,\gamma}$.}\label{fig:1}
\end{figure}

Then we study the probability density function (PDF) $P=P(x,y,z,t)$, which describes 
the 3D diffusion process in the 3D fractal mesh, which we call ``sponge'', 
see Figure~\ref{fig:1}. This fractal structure also defines the fractal structure 
of the nonzero diagonal components of the diffusion matrix $\hat{D}$, 
which are not zero only inside this fractal structure. That is,
 \begin{equation}\label{D-tensor}
    \hat{D}_{ij}= D_i\delta_{i,j}\sum_{\substack{l_{\beta}\in S_{\beta} \\ 
    l_{\gamma}\in S_{\gamma}}}
	\delta(y-l_{\beta})\delta(z-l_{\gamma}),   \quad   i,j=\{x,y,z\}.
\end{equation}
 
\section{Sponge diffusion equation}\label{sde}

The corresponding diffusion equation according to the 3D generalization 
of~\cite{sandev2017anomalous} reads as follows
\begin{multline}\label{spe-1}
	\partial_tP(x,y,z,t)=
	D_x\sum_{\substack{l_{\beta}\in S_{\beta} \\ l_{\gamma}\in S_{\gamma}}}
	\delta(y-l_{\beta})\delta(z-l_{\gamma})\partial_x^2 P(x,y,z,t) \\
	+D_y 	\sum_{\substack{l_{\alpha}\in S_{\alpha} \\ l_{\gamma}\in S_{\gamma}}} 
    \delta(x-l_{\alpha})\delta(z-l_{\gamma})\partial_y^2P(x,y,z,t) 
	+D_z \sum_{\substack{l_{\alpha}\in S_{\alpha} \\ l_{\beta}\in S_{\beta}}}
	\delta(x-l_{\alpha})\delta(y-l_{\beta})\partial_z^2P(x,y,z,t),
\end{multline}
where all variables and parameters are dimensionless.
In this case $D_i$, $i=\{x,y,z\}$ is a constant part of the diffusion coefficients 
in the corresponding direction; it is nonzero only inside the fractal structure, 
which is determined by a structure of the Dirac delta functions. Suppose transport 
is slow, when $D_i$ are small,
we can take the zero boundary conditions at infinity, while the initial 
condition $P(x,y,z,t=0)=P_0(x,y,z)$ is chosen at $t=0$ and its explicit form will 
be specified as necessary.

As mentioned above, we are interested in transport in the $x$ direction. To this end, 
we obtain the diffusion equation for the marginal PDF $P_1(x,t)$, 
which results from the integration of the PDF $P(x,y,z,t)$ with 
respect to (w.r.t.) $y$ and $z$ coordinates,
\begin{equation}\label{spe-2}
	P_1(x,t)=\int_{-\infty}^{\infty}dy\, P_2(x,y,t)=
    \int_{-\infty}^{\infty} dy\int_{-\infty}^{\infty}dz\, P(x,y,z,t)    .
\end{equation}
To perform these integrations of eq.~\eqref{spe-1}, we first present it in Laplace space by means of the Laplace transformation 
$\hat{P}(\dots,s)=\int_0^{\infty}P(\dots,t)e^{-st}dt$.
We also take into account that integration w.r.t. $x$ is not performed, then the summation
over the $x$ fractal set can be replaced by the fractal density, that is,
$\sum_{l_{\alpha}\in S_{\alpha}}\delta(x-l_{\alpha})\rightarrow\rho_{\alpha}(|x|)=
	|x|^{\alpha-1}/\Gamma(\alpha)$. All of these procedures eventually lead to
    the following result
\begin{align}\label{spe-3}
	s\hat{P}(x,y,z,s)-P_0(x,y,s)= &D_x\sum_{\substack{l_{\beta}\in S_{\beta} \\ 
    l_{\gamma}\in S_{\gamma}}}
	\delta(y-l_{\beta})\delta(z-l_{\gamma})\partial_x^2 \hat{P}(x,y,z,s)\nonumber\\&+
	D_y\rho_{\alpha}(|x|)
	\sum_{l_{\gamma}\in S_{\gamma}} \delta(z-l_{\gamma})\partial_y^2
    \hat{P} (x,y,z,s)\nonumber\\&
	+ D_z \rho_{\alpha}(|x|) \sum_{ l_{\beta}\in S_{\beta}}
    \delta(y-l_{\beta})\partial_z^2\hat{P}(x,y,z,s) .
\end{align}
Integrating eq.~\eqref{spe-3} w.r.t. $z$ and taking into account the boundary conditions at infinity, and eq.~\eqref{spe-2}, we obtain 
\begin{align}\label{spe-4}
	s\hat{P}_2(x,y,s)-P_{2,0}(x,y)= &D_x\sum_{\substack{l_{\beta}\in S_{\beta} \\ l_{\gamma}\in S_{\gamma}}}
	\delta(y-l_{\beta})\delta(z-l_{\gamma})\partial_x^2 \hat{P}(x,y,z=l_{\gamma},s)\nonumber\\&+
	D_y\rho_{\alpha}(|x|)
	\sum_{l_{\gamma}\in S_{\gamma}} \delta(z-l_{\gamma})\partial_y^2\hat{P} (x,y,z=l_{\gamma},s).
\end{align}
Let us represent the PDF $\hat{P}(x,y,z,s)$ in the form to compensate $s\hat{P}(x,y,z,s)$ in the l.h.s. of eq.~\eqref{spe-3}~\cite{ArBa91}, which reads\footnote{This ansatz is a generalization of a standard $2D$ comb consideration, namely it a map of a $2D$ comb consideration on the backbone description \cite{ArBa91,IoMeHo2018}. The $2D$ comb eq. \eqref{in-eq1}  
in Laplace space is 
$s\hat{P}-P_0=D_1\delta(y)\partial_x^2\hat{P}+D_2\partial_y^2P$.
For $y\neq 0$, the backbone term $D_1\delta(y)\partial_x^2\hat{P}$
disappears from the equation. Therefore, to consider diffusion on the backbone, 
one should compensate the terms related to diffusion in fingers with $y\neq 0$.
Taking into account that $2\delta(y)=d^2|y|/dy^2$, one present the ansatz
in the form $\hat{P}(x,y,s)=e^{-|y|\sqrt{s/D_2}}$. See also illuminated explanations in Refs. \cite{LZ98,IoZaPf16}.}

\begin{equation}\label{spe-5}
	\hat{P}(x,y,z,s)=f_2(x,y,s)\,\exp\left(-R^{1/2}_{\alpha}(x)\,s^{1/2}\,|z|\right),
\end{equation}
where $R_{\alpha}(x)=[D_z|x|^{\alpha-1}/\Gamma(\alpha)]^{-1}$.
This also yields the relation between $\hat{P}_2(x,y,s)$ and $f_2(x,y,s)$. Namely, performing integration w.r.t. $z$, from eq.~\eqref{spe-5}, we obtain
\begin{equation}\label{spe-6}
	\hat{P}_2(x,y,s)=\int_{-\infty}^{\infty} dz\, \hat{P}(x,y,z,s)=2\,f_2(x,y,s)\,R^{-1/2}_{\alpha}(x)\,s^{-1/2}.
\end{equation}
The summation in eq.~\eqref{spe-4} is over the fractal set $S_{\gamma}$, which
corresponds to integration over the fractal measure of the  
fractal density \eqref{density-1d}, $\rho_\gamma(|z|)=|z|^{\gamma-1}/\Gamma(\gamma)$   
~\cite{tar2004}.
Then, one finds
\begin{align}\label{spe-7}
	\sum_{l_{\gamma}\in S_{\gamma}}\hat{P}(x,y,z=l_{\gamma},s) &=f_2(x,y,s) \,
	2\int_0^{\infty}dl\,\frac{l^{\gamma-1}}{\Gamma(\gamma)}
		\exp\left(-R^{1/2}_{\alpha}(x)\,s^{1/2}\,l\right) \nonumber\\&
		=2\,f_2(x,y,s)\, \left[R^{1/2}_{\alpha}(x)\,s^{1/2}\right]^{-\gamma} 
		=\hat{P}_2(x,y,s)\left[R^{1/2}_{\alpha}(x)\,s^{1/2}\right]^{1-\gamma} ,
\end{align}
where we use eq.~\eqref{spe-6}. The next step is integration w.r.t. $y$ in eq.~\eqref{spe-4}, which after integration w.r.t. $z$ now reads as follows
\begin{align}\label{spe-8}
	s\hat{P}_2(x,y,s)-P_{2,0}(x,y)= &D_x\sum_{l_{\beta}\in S_{\beta} }
	\delta(y-l_{\beta})\partial_x^2 \left[R^{1/2}_{\alpha}(x)\,s^{1/2}\right]^{1-\gamma} \hat{P}_2(x,y,s) \nonumber\\&+
	D_y\rho_{\alpha}(|x|)\left[R^{1/2}_{\alpha}(x)\,s^{1/2}\right]^{1-\gamma}
	\partial_y^2 \hat{P}_2(x,y,s).
\end{align}
By analogy with eq.~\eqref{spe-5} the PDF $\hat{P}_2(x,y,s)$ can be represented as
\begin{equation}\label{spe-9}
	\hat{P}_2(x,y,s)=f_1(x,s)\,\exp\left(-R^{1/2}_{\beta}(x)\,s^{(\gamma+1)/4}\,|y|\right),
\end{equation}
which also yields the relation between $ f_1(x,s)$ and $\hat{P}_1(x,s)$,
\begin{equation}\label{spe-10}
	f_1(x,s)=\frac{1}{2}\left[R_{\beta}(x)\,s^{(\gamma+1)/2}\right]^{1/2}\hat{P}_1(x,s),
\end{equation}
where $R_{\beta}(x)=\left[ D_y\rho_{\alpha}(|x|)\right]^{-1}\left[R^{1/2}_{\alpha}(x)\right]^{\gamma-1}$.
Now, performing integration w.r.t. $y$ and then summation over the fractal set 
$S_{\beta}$ with the fractal density $\rho_\beta(|y|)=|y|^{\beta-1}/\Gamma(\beta)$,
one arrives at an expression analogous to eq.~\eqref{spe-7}, which reads
\begin{align}\label{spe-11}
	\sum_{l_{\beta}\in S_{\beta}}\hat{P}_2(x,y=l_{\beta},s) &=f_1(x,s) \cdot
	2\int_0^{\infty}dl\,\frac{l^{\beta-1}}{\Gamma(\beta)}
	\exp\left[-R^{1/2}_{\beta}(x)\,s^{(\gamma+1)/4}l\right]\nonumber\\& 
	=2 \left[R^{1/2}_{\beta}(x)\,s^{(\gamma+1)/4}\right]^{-\beta} f_1(x,s)
	=\left[R^{1/2}_{\beta}(x)\,{s}^{(\gamma+1)/4}\right]^{1-\beta} \hat{P}_1(x,s).
\end{align}
Eventually, we arrive at the desired equation for the Laplace transform $\hat{P}_1(x,s)$.
It reads as follows
\begin{equation}\label{spe-12}
	s\hat{P}_1(x,s)-P_{1,0}(x)= D_x\,s^{(3-\beta-\gamma-\beta\gamma)/4}
	\partial_x^2\left[ R^{(1-\gamma)/2}_{\alpha}(x)\, R^{(1-\beta)/2}_{\beta}(x) \hat{P}_1(x,s)\right].
\end{equation}
Performing the inverse Laplace transformation, we obtain the following generalized Fokker-Planck equation
\begin{equation}\label{spe-12-inv}
	\partial_tP_1(x,t)= D_x \frac{\partial}{\partial t}\int_{0}^{t}dt'\,\eta(t-t')\,
	\partial_x^2 \left[R^{(1-\gamma)/2}_{\alpha}(x)\, R^{(1-\beta)/2}_{\beta}(x)\, P_1(x,t')\right]
\end{equation}
with the memory kernel $\eta(t)=\mathcal{L}^{-1}\left[s^{-\mu}\right]=t^{\mu-1}/\Gamma(\mu)$, $\mu=(1+\beta+\gamma+\beta\gamma)/4$. In addition to classical diffusion problems, such memory effects also occur in constrained quantum motion in disordered media\footnote{A subordination approach to eq.~(\ref{spe-12-inv}) is considered in Appendix~\ref{app_sub}.}~\cite{petreska2020time,IoMeHo2018,lenzi2021fractional,sandev2025standard,trajanovski2026anomalous}. 

Equation~(\ref{spe-12-inv}) can be considered as a generalized Fokker-Planck equation with a memory kernel and a position-dependent diffusion coefficient $\mathcal{D}(x)\sim |x|^{(1-\alpha)(1-\mu)}$. More specifically, it is a heterogeneous Fokker-Planck equation in the It\^{o} interpretation, see, for example, Ref.~\cite{SaIo2022,srokowski2006diffusion,leibovich2019infinite}. Thus, the fractal structure of the fractal sponge induces a memory and heterogeneity in the system. 

It is worth noting that the same equation can be derived from the following coupled 
Langevin equations
\begin{align}\label{langevin}
    \left\lbrace\begin{array}{ll}
    \dot{x}(\tau)=\sqrt{2\mathcal{D}(x)}\,\zeta(\tau),  \\
    \dot{t}(\tau)=\xi(\tau), 
    \end{array}\right.
\end{align}
where $\zeta(\tau)$ is multiplicative white noise, and $\xi(\tau)$ is a $\mu$-stable L\'evy noise with L\'evy index in Laplace space given by $\hat{\Psi}(s)=1/\hat{\eta}(s)=s^{\mu}$. This is a subordinated heterogeneous diffusion process in which multiplicative noise is taken in the It\^{o} interpretation, see Appendix~\ref{app_sub} for details.

\section{Solution and transport characteristics}\label{solut}

After substitution $\hat{F}_{1}(x,s)=|x|^{\frac{(1-\alpha)(3-\beta-\gamma-\beta\gamma)}{4}}\hat{P}_1(x,s)$ in eq.~(\ref{spe-12}), we obtain the following equation
\begin{align}\label{spe-12-2}
	s|x|^{-\frac{(1-\alpha)(3-\beta-\gamma-\beta\gamma)}{4}}\hat{F}_1(x,s) 
    -&\frac{D_x}{D_y^{(1-\beta)/2}D_z^{(1-\gamma)(1+\beta)/4}} \nonumber\\
    &\times
    s^{(3-\beta-\gamma-\beta\gamma)/4}
	\Gamma(\alpha)^{(3-\beta-\gamma-\beta\gamma)/4}\partial_x^2  \hat{F}_1(x,s)=\delta(x),
\end{align}
where the initial condition is given by $P_{1,0}(x)=\delta(x)$. First, we solve the homogeneous equation
\begin{equation}\label{spe-12-2-h}
	\frac{D_y^{(1-\beta)/2}D_z^{(1-\gamma)(1+\beta)/4}}{D_x}\frac{s^{1-(3-\beta-\gamma-\beta\gamma)/4}}{\Gamma(\alpha)^{(3-\beta-\gamma-\beta\gamma)/4}}|x|^{-\frac{(1-\alpha)(3-\beta-\gamma-\beta\gamma)}{4}}\hat{G}_1(x,s)=
	\partial_x^2  \hat{G}_1(x,s),
\end{equation}
which is symmetric with respect to $x\rightarrow-x$. This is a Bessel-type (or Lommel-type) equation~\cite{GrRy07,JaEmLo60}
\begin{align}\label{lommel}
u''(y)+\frac{1-2\bar{\beta}}{y}u'(y)+\left[\left(a\bar{\alpha}y^{\bar{\alpha}-1}\right)^{2}+
\frac{\bar{\beta}^{2}-\nu^{2}\bar{\alpha}^{2}}{y^2}\right]u(y)=0,
\end{align}
where $a$, $\nu$, $\bar{\alpha}$, and $\bar{\beta}$ are parameters, while the prime symbol for $u$ denotes derivatives w.r.t. $y$ coordinate.\footnote{It should not be confused with the $y$ coordinate considered in the sponge diffusion equation in Sec. \ref{sde}.} The solution of eq.~(\ref{lommel}) is $$u(y)=y^{\bar{\beta}}Z_{\nu}\left(\imath a y^{\bar{\alpha}}\right),$$ where $Z_{\nu}(y)=C_{1}J_{\nu}(y)+C_{2}Y_{\nu}(y)$ is the Bessel function. For zero boundary conditions at infinity, the solution becomes \cite{SaIoMe16,sandev2017anomalous}
$$u(y)=y^{\bar{\beta}}K_{\nu}\left(a y^{\bar{\alpha}}\right),$$ where $K_{\nu}(y)$ is the modified Bessel function (of the third kind) \cite{AbSt72}. Therefore, the solution of eq.~(\ref{spe-12-2-h}) reads
\begin{align}\label{solK}
    \hat{G}_1(x,s)&=|x|^{1/2}K_{1/\bar{\alpha}}\left(\frac{2s^{\mu/2}|x|^{\bar{\alpha}/2}}{\bar{\alpha}}\sqrt{	\frac{D_y^{(1-\beta)/2}D_z^{(1-\gamma)(1+\beta)/4}}{D_x \Gamma(\alpha)^{(3-\beta-\gamma-\beta\gamma)/4}}}\right)\nonumber\\&=\frac{|x|^{1/2}}{2}H_{0,2}^{2,0}\left[\left.\frac{s^{\mu}|x|^{\bar{\alpha}}}{\bar{\alpha}^2}	\frac{D_y^{(1-\beta)/2}D_z^{(1-\gamma)(1+\beta)/4}}{D_x \Gamma(\alpha)^{(3-\beta-\gamma-\beta\gamma)/4}}\right|\begin{array}{cc}
      - \\
      \left(\frac{1}{2\bar{\alpha}},1\right), \left(-\frac{1}{2\bar{\alpha}},1\right)
    \end{array}\right],
\end{align}
where $\bar{\alpha}=\left[8-(1-\alpha)(3-\beta-\gamma-\beta\gamma)\right]/4=2-(1-\alpha)(1-\mu)$ and $\mu=(1+\beta+\gamma+\beta\gamma)/4$, while 
$H_{p,q}^{m,n}(z)$ is the Fox $H$-function~(\ref{H_integral}).
For the inhomogeneous equation, we use $\hat{F}_{1}(x,s)=C(s)\hat{G}_1(|x|,s)=C(s)\hat{G}_1(\xi,s)$, $|x|=\xi$. With this exchange of variables, the partial differentiations  
w.r.t. $x$ become
\begin{align}
    \frac{\partial}{\partial x}\hat{F}_1(\xi,s)=\frac{\partial}{\partial \xi}\hat{F}_1(\xi,s)\frac{d\xi}{dx}=\left[2\theta(x)-1\right]\frac{\partial}{\partial \xi}\hat{F}_1(\xi,s),
\end{align}
\begin{align}
    \frac{\partial^2}{\partial x^2}\hat{F}_1(\xi,s)&=\frac{\partial}{\partial x}\left\{\left[2\theta(x)-1\right]\frac{\partial}{\partial \xi}\hat{F}_1(\xi,s)\right\}=\left[2\theta(x)-1\right]^2\frac{\partial^2}{\partial\xi^2}\hat{F}_1(\xi,s)+2\delta(x)\frac{\partial}{\partial\xi}\hat{F}_1(\xi,s)\nonumber\\&=\frac{\partial^2}{\partial\xi^2}\hat{F}_1(\xi,s)+2\delta(x)\left.\frac{\partial}{\partial\xi}\hat{F}_1(\xi,s)\right|_{\xi=|x|=0}.
\end{align}
Substituting these results into eq.~(\ref{spe-12-2}) and collecting the terms containing the Dirac delta function $\delta(x)$, on both sides of the equation, we obtain
\begin{align}
    -2\left[	\frac{D_x \Gamma(\alpha)^{(3-\beta-\gamma-\beta\gamma)/4}}{D_y^{(1-\beta)/2}D_z^{(1-\gamma)(1+\beta)/4}}s^{1-\mu}\right]\frac{\partial}{\partial y}F_1(\xi=0,s)=1.
\end{align}
From this equation and using the series expansion of the modified Bessel function
\begin{align}
    K_{\nu}(z)\simeq \frac{\Gamma(\nu)}{2}\left(\frac{z}{2}\right)^{-\nu}\left[1+\frac{z^2}{4(1-\nu)}+\dots\right]+\frac{\Gamma(-\nu)}{2}\left(\frac{z}{2}\right)^{\nu}\left[1+\frac{z^2}{4(1+\nu)}+\dots\right], \,\, z\rightarrow0, \,\, \nu\notin Z, 
\end{align}
in eq.~(\ref{solK}), we find the constant $C(s)$. Eventually, the solution is
\begin{align}\label{pdf_spnge_x}
    P_{1}(x,t)=&\frac{\left(	\frac{D_y^{(1-\beta)/2}D_z^{(1-\gamma)(1+\beta)/4}}{D_x \Gamma(\alpha)^{1-\mu}}\right)^{1-\frac{1}{2\bar{\alpha}}}}{2\bar{\alpha}^{1-1/\bar{\alpha}}\Gamma(1-1/\bar{\alpha})}\frac{|x|^{\bar{\alpha}-3/2}}{t^{\left(1-\frac{1}{2\bar{\alpha}}\right)\mu}}\nonumber\\&\times H_{1,2}^{2,0}\left[\left.\frac{1}{\bar{\alpha}^2}	\frac{D_y^{(1-\beta)/2}D_z^{(1-\gamma)(1+\beta)/4}}{D_x \Gamma(\alpha)^{1-\mu}}\frac{|x|^{\bar{\alpha}}}{t^{\mu}}\right|\begin{array}{cc}
      \left(1-\mu+\frac{\mu}{2\bar{\alpha}},\mu\right) \\
      \left(\frac{1}{2\bar{\alpha}},1\right), \left(-\frac{1}{2\bar{\alpha}},1\right)
    \end{array}\right],
\end{align}
where we use the Laplace transformation formula~(\ref{Laplace H}) for the Fox $H$-function.

\begin{figure}[h!]
 \begin{center}
 \includegraphics[width=8cm]{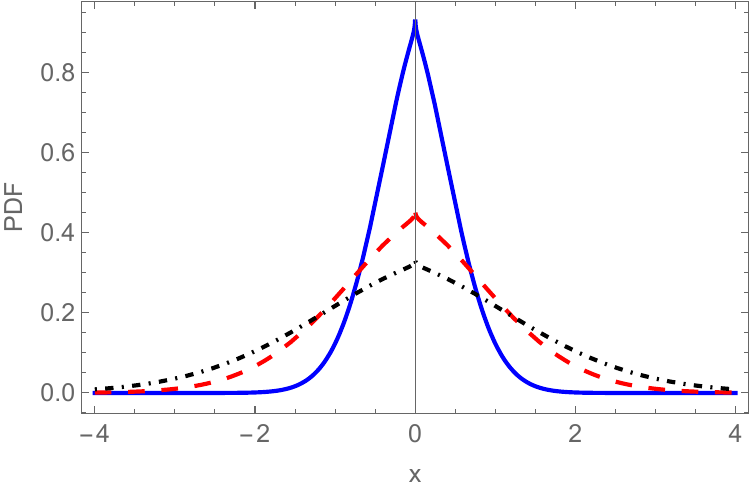}% This is a *.eps file
 \end{center}
 \caption{PDF~(\ref{pdf_spnge_x}) for $\alpha=\beta=\gamma=0.9$ and $t=0.1$ (blue solid line), $t=1$ (red dashed line) and $t=2$ (black dot-dashed line). We set $D_x=D_y=D_z=1$. }\label{fig:pdf1}
 \end{figure}

Using property (\ref{H_property-a}) and the Mellin transformation formula~(\ref{integral of H}), 
we obtain that the PDF is normalized, that is, $\int_{-\infty}^{\infty}P_{1}(x,t)\,dx=1$. The graphical representation of PDF~(\ref{pdf_spnge_x}) for the symmetric fractal sponge 
with $\alpha=\beta=\gamma$ is presented in Figure~\ref{fig:pdf1}.

Using relations~(\ref{H_property-a}) and ~(\ref{integral of H}), we find the MSD
\begin{align}\label{msd_sponge}
    \langle x^2(t)\rangle&=\frac{2\Gamma(2/\bar{\alpha})\Gamma(1/\bar{\alpha})\bar{\alpha}^{4/\bar{\alpha}-2}}{\Gamma(1-1/\bar{\alpha})}\left[\frac{D_x \Gamma(\alpha)^{1-\mu}}{D_y^{(1-\beta)/2}D_z^{(1-\gamma)(1+\beta)/4}}\right]^{2/\bar{\alpha}}\frac{t^{2\mu/\bar{\alpha}}}{\Gamma(1+2\mu/\bar{\alpha})}\nonumber\\&=2D_{\alpha,\beta,\gamma}t^{2\mu/\bar{\alpha}},
\end{align}
where (half of) the generalized diffusion coefficient is
\begin{align}\label{diff_coeff}    D_{\alpha,\beta,\gamma}=\frac{\Gamma(2/\bar{\alpha})\Gamma(1/\bar{\alpha})\bar{\alpha}^{4/\bar{\alpha}-2}}{\Gamma(1-1/\bar{\alpha})\Gamma(1+2\mu/\bar{\alpha})}\left[\frac{D_x \Gamma(\alpha)^{1-\mu}}{D_y^{(1-\beta)/2}D_z^{(1-\gamma)(1+\beta)/4}}\right]^{2/\bar{\alpha}}.
\end{align}
Thus, the MSD has the following behavior
\begin{align}
    \langle x^2(t)\rangle\simeq t^{\frac{2(1+\beta+\gamma+\beta\gamma)}{8-(1-\alpha)(3-\beta-\gamma-\beta\gamma)}},
\end{align}
which means subdiffusion since the transport exponent is
$\frac{1}{4}\leq\frac{2(1+\beta+\gamma+\beta\gamma)}{8-(1-\alpha)(3-\beta-\gamma-\beta\gamma)}\leq1$. 

\begin{figure}[h!]
 \begin{center}
 \includegraphics[width=8cm]{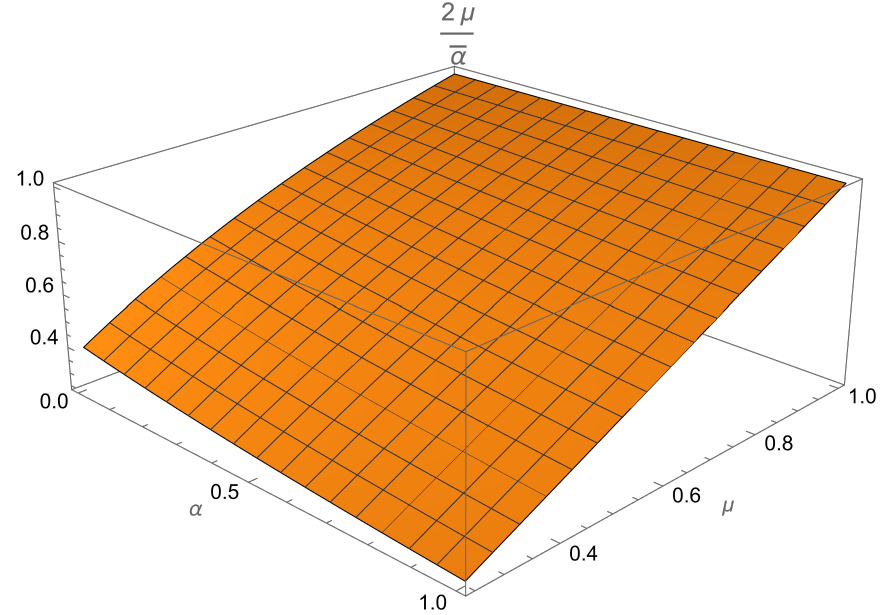}% This is a *.eps file
 \end{center}
 \caption{Dependence of transport exponent $2\mu/\bar{\alpha}$ on $\mu$ and $\alpha$. }\label{fig:tr1}
 \end{figure}

\begin{figure}[h!]
 \begin{center}
 \includegraphics[width=8cm]{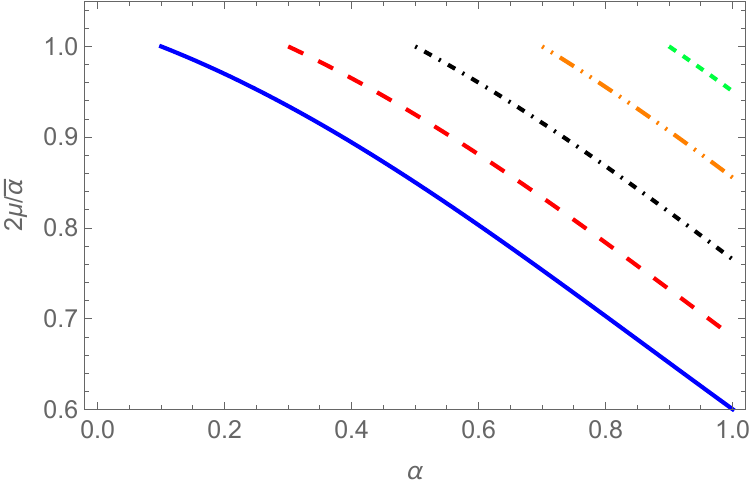}% This is a *.eps file
 \end{center}
 \caption{Dependence of transport exponent $2\mu/\bar{\alpha}$ on $\alpha$ for $d_f=2.1$ (blue solid line); $d_f=2.3$ (red dashed line); $d_f=2.5$ (black dot-dashed line); $d_f=2.7$ (orange dot-dot-dashed line); $d_f=2.9$ (dotted green line). Here we use  $d_f=\alpha+\beta+\gamma$, with $\beta=\gamma$ and $D_x=D_y=D_z=1$. }\label{fig:tr2}
 \end{figure}

\begin{figure}[h!]
 \begin{center}
 \includegraphics[width=8cm]{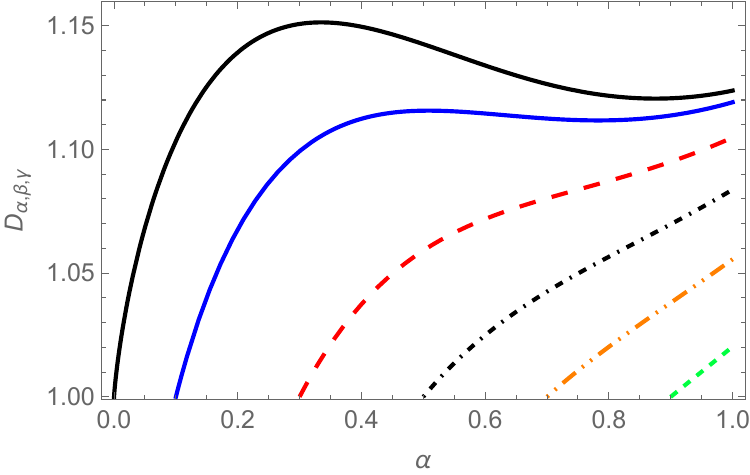} \includegraphics[width=8cm]{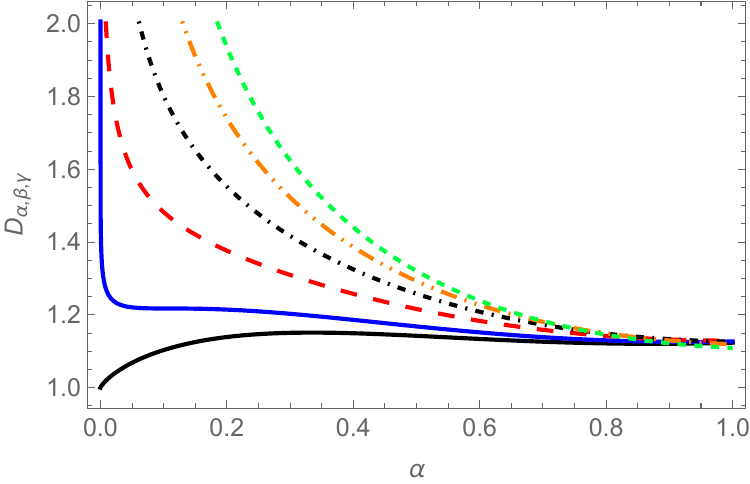}% This is a *.eps file
 \end{center}
 \caption{Dependence of $D_{\alpha,\beta,\gamma}$, eq.~(\ref{diff_coeff}), on $\alpha$; (left panel with $2\leq d_f<3$) $d_f=2.0$ (black solid line), $d_f=2.1$ (blue solid line), $d_f=2.3$ (red dashed line), $d_f=2.5$ (black dot-dashed line), $d_f=2.7$ (orange dot-dot-dashed line), $d_f=2.9$ (dotted green line); (right panel with $1<d_f\leq 2$) $d_f=2.0$ (black solid line), $d_f=1.9$ (blue solid line), $d_f=1.7$ (red dashed line), $d_f=1.5$ (black dot-dashed line), $d_f=1.3$ (orange dot-dot-dashed line), $d_f=1.1$ (dotted green line). Black solid lines in both panels correspond to $d_f=2$.
 Here we use  $d_f=\alpha+\beta+\gamma$ with $\beta=\gamma$ and $D_x=D_y=D_z=1$. }\label{fig:D}
 \end{figure}

The behavior of the transport exponent $2\mu/\bar{\alpha}=2\mu\left[2-(1-\alpha)(1-\mu)\right]^{-1}$ as a function of 
$\mu=(1+\beta+\gamma+\beta\gamma)/4$ and $\alpha$ is given in Figure~\ref{fig:tr1}. 
It describes the transport exponent for all possible realizations of the fractional dimension
$d_f=\alpha+\beta+\gamma$ of the sponge. 
Some specific realizations of the transport exponent for some
fixed values of the fractal dimension $d_f$ of the sponge are defined by the plots in 
Figure ~\ref{fig:tr2}.
The dependence of the generalized diffusion coefficient $D_{\alpha,\beta,\gamma}$
on the fractal dimension $\alpha$ for fixed $d_f$ is shown in Figure~\ref{fig:D}.

% \begin{figure}[h!]
%  \begin{center}
%  \includegraphics[width=11cm]{DvsAlpha.pdf}% This is a *.eps file
%  \end{center}
%  \caption{Dependence of the diffusion coefficient~(\ref{diff_coeff}) on $\alpha$ for $d_f=\alpha+\beta+\gamma=2.5$ and $\beta=\gamma$. We set $D_x=D_y=D_z=1$. }\label{fig:D}
%  \end{figure}

% \begin{figure}[h!]
%  \begin{center}
%  \includegraphics[width=11cm]{DvsAlpha2.pdf}% This is a *.eps file
%  \end{center}
%  \caption{Dependence of the diffusion coefficient~(\ref{diff_coeff}) on $\alpha$ for $d_f=\alpha+\beta+\gamma=2.5$ and $\beta=\gamma$. We set $D_x=D_y=D_z=1$. {\bf Here in the diffusion coefficient I also take $1/\Gamma(1+2\mu/\bar{\alpha})$.}}\label{fig:D}
%  \end{figure}

\subsection{Limiting cases and asymptotic behavior}

Here, we will analyze the asymptotic behavior of the PDF. Let us rewrite the PDF as
\begin{align}\label{pdf_spnge_x2}
    P_{1}(x,t)=\frac{|x|^{\bar{\alpha}-3/2}}{2\bar{\alpha}^{1-1/\bar{\alpha}}\Gamma(1-1/\bar{\alpha})\left[Dt^{\mu}\right]^{\left(1-\frac{1}{2\bar{\alpha}}\right)}} H_{1,2}^{2,0}\left[\left.\frac{|x|^{\bar{\alpha}}}{\bar{\alpha}^2Dt^{\mu}}\right|\begin{array}{cc}
      \left(1-\mu+\frac{\mu}{2\bar{\alpha}},\mu\right) \\
      \left(\frac{1}{2\bar{\alpha}},1\right), \left(-\frac{1}{2\bar{\alpha}},1\right)
    \end{array}\right],
\end{align}
where $D=D_x \Gamma(\alpha)^{1-\mu}\left[D_y^{(1-\beta)/2}D_z^{(1-\gamma)(1+\beta)/4}\right]^{-1}$. For $|x|^{\bar{\alpha}}\left[\bar{\alpha}^2Dt^{\mu}\right]^{-1}\gg1$, using the asymptotic formula (\ref{H_asymptotic}), we obtain the following behavior
\begin{align}\label{pdf_spnge_x2asympt}
    P_{1}(x,t)\sim &\frac{\mu^{(\mu-\frac{\mu}{2\bar{\alpha}}-1/2)(\frac{\mu}{2-\mu}+1)-\frac{\mu}{2(2-\mu)}}(2-\mu)^{-1/2}}{2\bar{\alpha}^{1-1/\bar{\alpha}+2\frac{\mu-\frac{\mu}{2\bar{\alpha}}-1}{2-\mu}}\Gamma(1-1/\bar{\alpha})}\frac{|x|^{\frac{\bar{\alpha}-\mu/2}{2-\mu}-3/2}}{\left[Dt^{\mu}\right]^{\frac{1}{2-\mu}\left(1-\frac{1}{\bar{\alpha}}\right)}} \nonumber\\&\times \exp\left([2-\mu]\mu^{\frac{\mu}{2-\mu}}\left[\frac{|x|^{\bar{\alpha}}}{\bar{\alpha}^2Dt^{\mu}}\right]^{\frac{1}{2-\mu}}\right). 
\end{align} 
The asymptotic behavior in eq. \eqref{pdf_spnge_x2asympt}
is valid for all time scales, including  a short-time scale, when 
$t\ll\left[|x|^{\bar{\alpha}}/\left(\bar{\alpha}^2D\right)\right]^{1/\mu}$. In this case, this
explicit form can also be convenient for the interpretation of experimental data
\cite{baklanov2026}. 

In the opposite case $|x|^{\bar{\alpha}}\left[\bar{\alpha}^2Dt^{\mu}\right]^{-1}\ll1$, which also means $t\gg\left[|x|^{\bar{\alpha}}/\left(\bar{\alpha}^2D\right)\right]^{1/\mu}$, the series expansion 
of the Fox $H$-function~(\ref{H_expansion}) yields
\begin{align}
    P_1(x,t)\sim \frac{\bar{\alpha}\Gamma(1/\bar{\alpha})}{\Gamma(1-1/\bar{\alpha})\Gamma(1-\mu+\mu/\bar{\alpha})}\frac{|x|^{\bar{\alpha}-2}}{\left(Kt^\mu\right)^{1-1/\bar{\alpha}}},
\end{align}
where $K=\left[\bar{\alpha}^{-2}	D_y^{(1-\beta)/2}D_z^{(1-\gamma)(1+\beta)/4}\left/\left(D_x \Gamma(\alpha)^{1-\mu}\right)\right.\right]^{-1}$. For $\alpha=\beta=\gamma=1$, which means $\bar{\alpha}=2$ and $\mu=1$, the result reduces to $P_1(x,t)\sim 1\left/\sqrt{4\pi D_x t}\right.$, as expected for the standard diffusion process.
% \begin{align}\label{alpha}
% \alpha=2-\mu+\frac{\mu}{2\bar{\alpha}},
% \end{align}
% \begin{align}\label{m}
% m^{*}=2-\mu,
% \end{align}
% \begin{align}\label{C}
% C=\mu^\mu,
% \end{align}
% \begin{align}\label{B}
% B=(2\pi)^{0}\mu^{\mu(\mu-\frac{\mu}{2\bar{\alpha}}-1)/(2-\mu)}\left(2-\mu\right)^{-1/2}\mu^{\mu-\frac{\mu}{2\bar{\alpha}}-1/2}=\mu^{(\mu-\frac{\mu}{2\bar{\alpha}}-1/2)(\frac{\mu}{2-\mu}+1)-\frac{\mu}{2(2-\mu)}}(2-\mu)^{-1/2}.
% \end{align}

In general case of the argument $|x|^{\bar{\alpha}}\left[\bar{\alpha}^2Dt^{\mu}\right]^{-1}$, for $\alpha=\beta=\gamma=1$, the PDF~(\ref{pdf_spnge_x}) becomes Gaussian,
\begin{align}\label{gaussian}
    P_{1}(x,t)&=\frac{1}{\sqrt{4\pi D_x t}}\left(\frac{x^2}{4D_x t}\right)^{1/4} H_{0,1}^{1,0}\left[\left.\frac{x^2}{4D_x t}\right|\begin{array}{cc}
      - \\
      (-1/4,1)
    \end{array}\right]\nonumber\\&=\frac{1}{\sqrt{4\pi D_x t}} H_{0,1}^{1,0}\left[\left.\frac{x^2}{4D_x t}\right|\begin{array}{cc}
      - \\
      (0,1)
    \end{array}\right]=\frac{1}{\sqrt{4\pi D_x t}}e^{-\frac{x^2}{4D_xt}},
\end{align}
where we use property~(\ref{H_property2}) and relation~(\ref{Hexp}). In such a case, the MSD~(\ref{msd_sponge}) corresponds to normal diffusion, i.e., 
\begin{align}
    \langle x^2(t)\rangle=2D_xt.
\end{align}

% {\bf I should carefully check all the calculations!

% The norm should also be checked!}

% {\it \color{blue} Dear Trifce, another very important characteristic here is the transport constant as a function of fractal dimensions ! Probably with the same scaling it is possible to find this constant.}

% When $\alpha=\beta=\gamma=1$, normal diffusion takes place with $2\mu/\bar{\alpha}=1$.

\begin{figure}[h!]
 \begin{center}
 \includegraphics[width=8cm]{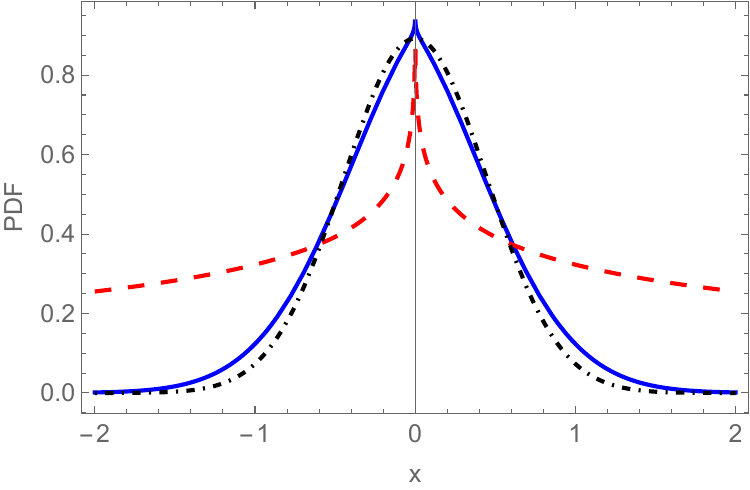}% This is a *.eps file
 \end{center}
 \caption{Comparison of PDF~(\ref{pdf_spnge_x}) for $\alpha=\beta=\gamma=0.9$ (blue solid line), PDF~(\ref{pdf_spnge_x_special}) ($\alpha=1$, $\beta=\gamma=0$) (red dashed line) and the Gaussian PDF~(\ref{gaussian}) ($\alpha=\beta=\gamma=1$) (black dot-dashed line) for $t=0.1$. We set $D_x=D_y=D_z=1$. The absence of singularity at $x=0$ of the Gaussian PDF should be pointed out.}\label{fig:pdf2}
 \end{figure}

In the case where $\alpha=1$, $\beta=\gamma=0$, the PDF becomes
\begin{align}\label{pdf_spnge_x_special}
    P_{1}(x,t)=&\frac{\left(	\frac{D_y^{1/2}D_z^{1/4}}{D_x}\right)^{3/4}}{\sqrt{8\pi}}\frac{|x|^{1/2}}{t^{3/16}}H_{1,2}^{2,0}\left[\left.	\frac{D_y^{1/2}D_z^{1/4}}{4D_x}\frac{|x|^{2}}{t^{1/4}}\right|\begin{array}{cc}
      \left(13/16,1/4\right) \\
      \left(1/4,1\right), \left(-1/4,1\right)
    \end{array}\right]\nonumber\\&=\frac{1}{2|x|^{1/4}}H_{1,2}^{2,0}\left[\left.	\sqrt{\frac{D_y^{1/2}D_z^{1/4}}{4D_x}}\frac{|x|^{1/4}}{t^{1/8}}\right|\begin{array}{cc}
      \left(1,1/8\right) \\
      \left(1,1/2\right), \left(1/2,1/2\right)
    \end{array}\right].
\end{align}
Subdiffusion with the minimal transport exponent $2\mu/\bar{\alpha}=1/4$ is reached
in this inhomogeneous sponge, that is,
\begin{align}
    \langle x^2(t)\rangle=2\frac{D_x}{\sqrt{D_y\sqrt{D_z}}}\frac{t^{1/4}}{\Gamma(5/4)}.
\end{align}
In this way, it becomes the 3D comb subdiffusion,  as the latter is also anisotropic \cite{DoMa-PuSaMeIoKo2020}. 
Different PDF realizations in eqs. (\ref{pdf_spnge_x}), (\ref{pdf_spnge_x_special})  and~(\ref{gaussian}) are given in Figure~\ref{fig:pdf2}. 

It should be pointed out that due to the symmetrical (isotropy) property of Eq. \eqref{spe-1},
the same result should be obtained for the MSD in the $y$  and $z$ directions. For example,
performing the same integrations w.r.t. $x$ and $z$, we obtain 
\begin{align}
    \langle y^2(t)\rangle\simeq t^{2\mu/\bar{\alpha}}=
    t^{\frac{2(1+\alpha+\gamma+\alpha\gamma)}{8-(1-\beta)(3-\alpha-\gamma-\alpha\gamma)}},
\end{align}
where we just replace $\alpha\leftrightarrow \beta$.

\section{Summary}\label{sum}

Considering anomalous transport in fractal, porous media, in essence, one meets two problems.  Listed in order, we have solved them simultaneously.
To study the idealized fractal sponge shown in Figure~2, our main intentions 
were (i) to discuss how fractional calculus can explain anomalous transport, including 
transport characteristics, in fractal porous media and (ii) to show how this 
fractional theory emergencies due to fractal geometry.
The fractal geometry of the sponge $S=S_{\alpha}\times S_{\beta}\times S_{\gamma}$ determines the matrix of diffusion coefficients \eqref{D-tensor} that eventually determines the Fokker-Planck equation \eqref{spe-1}, which determines the heat transport, described by the probability density function (PDF) $P(x,y,z)$. Following the comb strategy, a main direction
of transport has been chosen, namely the $x$-coordinate, which is described by the marginal PDF
\eqref{spe-2}, $P_1(x,t)=
    \int_{-\infty}^{\infty} dy\int_{-\infty}^{\infty}dz\, P(x,y,z,t)$.
The latter is governed by the time fractional Fokker-Planck equation (FFPE) 
\begin{equation}\label{sum-1}
	\partial_tP_1(x,t)= {}^{\rm RL}\partial^{\mu}_t\,	\partial_x^2 \, \mathcal{D}(x)
    P_1(x,t),
\end{equation}
where ${}^{\rm RL}\partial^{\mu}_t$ is the Riemann-Liouville fractional derivative
\cite{SaKiMa93,OlSp74,MiRo93,Po99},
\[
{}^{\rm RL}\partial^{\mu}_tf(t)=\frac{1}{\Gamma(1-\mu)}\frac{d}{dt}
\int_0^t(t-t')^{-\mu}f(t')dt'
\]
with $\mu=(1+\beta+\gamma+\beta\gamma)/4$, while the 
position-dependent diffusion coefficient is also a function of the fractal dimensions, 
$\mathcal{D}(x)\sim |x|^{(1-\alpha)(1-\mu)}$. The interplay of both the memory 
effect due to the Riemann-Liouville fractional derivative and the diffusion coefficient 
eventually results in subdiffusion with the transport exponent $1/4 \leq 2\mu/\bar{\alpha}\leq 1$, see eq. \eqref{msd_sponge}.
This fractional equation is universal and valid for the $y$ and $z$ directions by replacing
$\alpha\rightarrow\beta$ and $\alpha\rightarrow\gamma$, respectively.
That is, on the way of coarsening the description, the fractional diffusion equation 
\eqref{sum-1} emerges, reflecting the fractal geometry of the sponge, leading
to memory effects due to trapping.

The exact solution for the marginal PDF is obtained in the form of the Fox $H$-function
\eqref{pdf_spnge_x}. In this way, the MSD is also obtained together
with exact expressions of the transport exponent and the generalized diffusion 
coefficient as functions of fractal parameters $\alpha,\beta$ and $\gamma$. All possible 
realizations of the transport exponent for all possible realizations 
of $\alpha, \beta$ and $\gamma$ form a two dimensional surface depicted in 
Figure~\ref{fig:tr1}.
For the constraint condition in the form of the fractal dimension 
$d_f=\alpha+\beta+\gamma$, this surface degenerates into one-dimensional 
graphs, shown in Figure~\ref{fig:tr2}.
It should be noted that the fractional dimension $d_f$ is a macroscopic characteristic 
of the artificial porous media that can be obtained. For example, the fractal 
dimension of the low-$k$ dielectric is calculated in the framework of the 
Frenkel-Halsey-Hill model~\cite{PoTs2009,baklanov2026}.
In the same way of the topological and symmetrical constraints  $d_f=\alpha+\beta+\gamma$ and 
$\beta=\gamma$, the generalized diffusion coefficient depends on $\alpha=\alpha(d_f)$.
The resulting plots are depicted in Figure~\ref{fig:D},
which shows different functional  behavior of the generalized diffusion coefficient
as a function of $d_f$ in the left and right panels.
In the left panel with $d_f\geq 2$, $D_{\alpha,\beta,\gamma}$ is an increasing function at the
limits $\alpha\rightarrow 0$ and $\alpha\rightarrow 1$,
while in the right panel with $d_f<2$  it is decreasing function. 
The  plot with $d_f=2$ (solid black line) describes mixed behavior and separates 
these regions.

It should be admitted that this macroscopic FFPE is also supported by the 
microscopic phenomenon  
\cite{fogedby94,MeSt2013}  described in the framework of the
subordinated Langevin equation of the form of eq.~\eqref{langevin}, where
\begin{subequations}\label{sum-2}
\begin{align}
   & P_1(X,t)=\int_0^{\infty}f(X,\tau)h(\tau,t)d\tau, \quad X=x,y,z , \label{sum-2a} \\
   & f(X,\tau)=\langle\delta(X-X(\tau))\rangle_{\zeta},\quad 
   h(\tau,t) =-\partial_{\tau}\langle\left[\Theta(t-\xi(\tau)\right] \rangle_{\xi}  ,         \label{sum-2b}  
\end{align}
\end{subequations}
where $f(X,\tau)$ is the solution of eq.~\eqref{langevin} without memory kernel ($\eta(t)=1$), $\Theta(Y)$ is the Heaviside function and $\langle\dots\rangle$ defines the averaging
w.r.t. the corresponding random process $\xi$ or $\zeta$.  
Here $h(\tau,t)$ is a subordination PDF
such  that $\hat{h}(\tau,s)=\left[s\hat{\eta}(s)\right]^{-1}e^{-\tau/\hat{\eta}(s)}=s^{\mu-1}e^{-\tau s^\mu}$ is the $\mu$-stable L\'evy PDF~\cite{schneider1986stochastic}, given by
 \begin{align}
     h(\tau,t)=\mathcal{L}^{-1}\left[s^{\mu-1}e^{-\tau s^\mu}\right]=\frac{t}{\mu\tau^{1+1/\mu}}L_{\mu}\left(\frac{t}{\tau^{1/\mu}}\right).
 \end{align}

In conclusion, anomalous diffusion in fractal porous media is considered.
A model of a diffusive process inside  a fractal sponge structure is suggested.
The sponge is considered in the form of the direct product of Cantor sets 
$S=S_{\alpha}\times S_{\beta}\times S_{\gamma}$.
The  corresponding one-dimensional diffusive process is obtained and shown to be 
governed by a generalized Fokker-Planck equation with a power-law memory kernel 
and a position-dependent diffusion coefficient. 
That is, the fractal structure of the medium induces memory
effects and heterogeneity in the transport system. 
The considered model may be of interest to describe anomalous heat transport 
in porous fractal media, for example, in porous low-$k$ dielectric composites.

\appendix

\section{Subordinated heterogeneous diffusion process}\label{app_sub}

Consider the heterogeneous diffusion equation with memory kernel, see eq.~(\ref{spe-12-inv}),
\begin{equation}\label{spe-12-inv_app}
	\partial_tP_1(x,t)= D_x \frac{\partial}{\partial t}\int_{0}^{t}dt'\,\eta(t-t')\,
	\partial_x^2 \left[R^{(1-\gamma)/2}_{\alpha}(x)\, R^{(1-\beta)/2}_{\beta}(x)\, P_1(x,t')\right].
\end{equation}
By the Laplace transformation, one finds
\begin{equation}\label{spe-12_app}
	s\hat{P}_1(x,s)-\delta(x)= D_x\,s\hat{\eta}(s)\,
	\partial_x^2\left[ R^{(1-\gamma)/2}_{\alpha}(x)\, R^{(1-\beta)/2}_{\beta}(x) \hat{P}_1(x,s)\right].
\end{equation}
The corresponding equation for PDF $P_{0}(x,t)$ in the absence of the memory kernel
(namely $\eta(t)=1$, i.e., $\hat{\eta}(s)=1/s$) is
\begin{equation}\label{spe-12-inv_app00}
	\partial_tP_{0}(x,t)= D_x 
	\partial_x^2 \left[R^{(1-\gamma)/2}_{\alpha}(x)\, R^{(1-\beta)/2}_{\beta}(x)\, P_{0}(x,t)\right].
\end{equation}
In Laplace space, it reads
\begin{equation}\label{spe-12_app0}
	s\hat{P}_0(x,s)-\delta(x)= D_x\,
	\partial_x^2\left[ R^{(1-\gamma)/2}_{\alpha}(x)\, R^{(1-\beta)/2}_{\beta}(x) \hat{P}_0(x,s)\right].
\end{equation}
We introduce the substitution $s\rightarrow1/\hat{\eta}(s)$ in eq.~(\ref{spe-12_app0}), which yields
\begin{equation}\label{spe-12_app02}
	\frac{1}{\hat{\eta}(s)}\hat{P}_0(x,1/\hat{\eta}(s))-\delta(x)= D_x\,
	\partial_x^2\left[ R^{(1-\gamma)/2}_{\alpha}(x)\, R^{(1-\beta)/2}_{\beta}(x) \hat{P}_0(x,1/\hat{\eta}(s))\right].
\end{equation}
Then, introducing a new PDF
\begin{equation}
    \hat{P}_{sub}(x,s)=\frac{1}{s\hat{\eta}(s)}\hat{P}_{0}(x,1/\hat{\eta}(s))
\end{equation}
in eq.~(\ref{spe-12_app02}), we obtain
\begin{equation}\label{spe-12_app023}
	s\hat{P}_{sub}(x,s)-\delta(x)= D_x\,s\hat{\eta}(s)\,
	\partial_x^2\left[ R^{(1-\gamma)/2}_{\alpha}(x)\, R^{(1-\beta)/2}_{\beta}(x) \hat{P}_{sub}(x,s)\right].
\end{equation}
Comparing it with eq.~(\ref{spe-12_app}), we conclude that both equations are identical. That is, the solution of eq.~(\ref{spe-12-inv_app}) can be given in terms of the solution of eq.~(\ref{spe-12-inv_app00}), i.e., via PDF $P_{0}(x,t)$. Thus,
\begin{equation}
    \hat{P}_{1}(x,s)=\frac{1}{s\hat{\eta}(s)}\hat{P}_{0}(x,1/\hat{\eta}(s)).
\end{equation}
The inverse Laplace transformation yields the following subordination integral~\cite{metzler2000random,barkai2001fractional}
\begin{equation}
    P_{1}(x,t)=\int_{0}^{\infty}P_{0}(x,u)h(u,t)\,du,
\end{equation}
where $h(u,t)$ is a so-called subordination function, which has the form
\begin{align}
    h(u,t)=\mathcal{L}^{-1}[\hat{h}(u,s)]=\mathcal{L}^{-1}\left[\frac{1}{s\hat{\eta}(s)}e^{-u/\hat{\eta}(s)}\right].
\end{align}
The heterogeneous diffusion equation~(\ref{spe-12-inv_app00}) can be described in terms of the Langevin equation with position dependent diffusion coefficient $\mathcal{D}(x)=R^{(1-\gamma)/2}_{\alpha}(x)\, R^{(1-\beta)/2}_{\beta}(x)$ in It\^{o} interpretation, i.e.~\cite{leibovich2019infinite}
\begin{align}
    \dot{x}(t)=\sqrt{2\mathcal{D}(x)}\,\zeta(t),
\end{align}
where $\zeta(t)$ is a multiplicative white noise. 
Then the heterogeneous diffusion process with memory governed by eq.~(\ref{spe-12-inv_app}) and being the subordinated heterogeneous diffusion process, can be given in terms of the following coupled Langevin equation~\cite{fogedby94}
\begin{align}\label{langevin_app}
    \left\lbrace\begin{array}{ll}
    \dot{x}(u)=\sqrt{2\mathcal{D}(x)}\,\zeta(u),  \\
    \dot{t}(u)=\xi(u), 
    \end{array}\right.
\end{align}
where $\zeta(u)$ is multiplicative white noise, while $\xi(u)$ is a stable L\'evy noise with L\'evy index in Laplace space given by $\hat{\Psi}(s)=1/\hat{\eta}(s)$, and $u$ is the operational time, which is related to the physical time $t$ as follows $t=\int_0^u\xi(u')\,du'$.

\section{Fox $H$-function}

The Fox $H$-function is defined by means of the following Mellin-Barnes integral~\cite{MaHa08,MaSaHa10}
\begin{align}\label{H_integral}
H_{p,q}^{m,n}(z)=H_{p,q}^{m,n}\left[z\left|\begin{array}{c c}
    (a_1,A_1),...,(a_p,A_p)\\
    (b_1,B_1),...,(b_q,B_q)
  \end{array}\right.\right]=H_{p,q}^{m,n}\left[z\left|\begin{array}{c l}
    (a_p,A_p)\\
    (b_q,B_q)
  \end{array}\right.\right]=\frac{1}{2\pi\imath}\int_{\Omega}ds\,\theta(s)z^{-s},
\end{align}
where
\begin{align}\label{theta_H}
    \theta(s)=\frac{\prod_{j=1}^{m}\Gamma(b_j+B_js)\prod_{j=1}^{n}
    \Gamma(1-a_j-A_js)}{\prod_{j=m+1}^{q}\Gamma(1-b_j-B_js)\prod_{j=n+1}^{p}\Gamma(a_j+A_js)},
\end{align}
$0\leq n\leq p$, $1\leq m\leq q$, $a_i,b_j \in \mathrm{C}$, $A_i,B_j\in\mathrm{R}^{+}$, $i=1,...,p$, $j=1,...,q$. The contour integration $\Omega$ starts at $c-\imath\infty$ and finishes at $c+\imath\infty$ separating the poles of the function $\Gamma(b_j+B_js)$, $j=1,...,m$ from those of the function $\Gamma(1-a_i-A_is)$, $i=1,...,n$.

For $\delta>0$, the following property holds true
\begin{align}\label{H_property-a}
H_{p,q}^{m,n}\left[z^{\delta}\left|\begin{array}{l}
    (a_p,A_p)\\
    (b_q,B_q)
  \end{array}\right.\right]=\frac{1}{\delta} H_{p,q}^{m,n}\left[z\left|\begin{array}{c l}
    (a_p,A_p/\delta)\\
    (b_q,B_q/\delta)
  \end{array}\right.\right].
\end{align}
The Fox $H$-function has the property
\begin{align}\label{H_property2}
z^{\sigma}H_{p,q}^{m,n}\left[z\left|\begin{array}{c l}
    (a_p,A_p)\\
    (b_q,B_q)
  \end{array}\right.\right]=H_{p,q}^{m,n}\left[z\left|\begin{array}{c l}
    (a_p+\sigma A_p,A_p)\\
    (b_q+\sigma B_q,B_q)
  \end{array}\right.\right].
\end{align}

The exponential function is a special case of the Fox $H$-function
\begin{align}\label{Hexp}
H_{0,1}^{1,0}\left[z\left|\begin{array}{cc}
\\
(0,1)\end{array}\right.\right]=e^{-z}.
\end{align}

The Mellin transform of the Fox $H$-function is
\begin{align}\label{integral of H}
\int_{0}^{\infty}dx\,x^{\xi-1}H_{p,q}^{m,n}\left[ax\left|\begin{array}{l}
    (a_1,A_1),...,(a_p,A_p)\\
    (b_1,B_1),...,(b_q,B_q)
  \end{array}\right.\right]=a^{-\xi}\,\theta(\xi),
\end{align}
where
$$\theta(\xi)=\frac{\prod_{j=1}^{m}\Gamma(b_j+B_j\xi)\prod_{j=1}^{n}\Gamma(1-a_j-A_j\xi)}{\prod_{j=m+1}^{q}\Gamma(1-b_j-B_j\xi)\prod_{j=n+1}^{p}\Gamma(a_j+A_j\xi)}$$ is defined in eq.~\eqref{theta_H}.

The Laplace transformation formula for the Fox $H$-function is given by~\cite{MaHa08,MaSaHa10}
\begin{align}\label{Laplace H}
\mathcal{L}^{-1}\left[s^{-\rho}H_{p,q}^{m,n}\left[zs^{\sigma}\left|\begin{array}{c
l}
    (a_p,A_p)\\
    (b_q,B_q)
  \end{array}\right.\right]\right]
  =t^{\rho-1}H_{p+1,q}^{m,n}\left[zt^{-\sigma}\left|\begin{array}{c
l}
(a_p,A_p), (\rho,\sigma)\\
    (b_q,B_q)
  \end{array}\right.\right],
\end{align}
where $\rho,z,s\in \mathrm{C}$, $\Re(s)>0$, $\sigma>0$, $\Re(\rho)+\sigma\max_{1\le i\le n}\left[\frac{1}{A_i}-\frac{\Re(a_i)}{A_i}\right]>0$, $|\arg z|<\frac{\pi\theta}{2}$, $\theta=\alpha-\sigma$.

The asymptotic expansion of the Fox $H$-function $H_{p,q}^{m,0}(z)$ for large $z$ is \cite{schneider1989fractional}
\begin{align}\label{H_asymptotic}
H_{p,q}^{m,0}(z)\sim
Bz^{(1-\alpha)/m^{*}}\exp\left(-m^{*}C^{1/m^{*}}z^{1/m^{*}}\right),
\end{align}
where
\begin{align}\label{alpha}
\alpha=\sum_{k=1}^{p}a_{k}-\sum_{k=1}^{q}b_{k}+\frac{1}{2}(q-p+1),
\end{align}
\begin{align}\label{m}
m^{*}=\sum_{j=1}^{q}B_{j}-\sum_{j=1}^{p}A_{j}>0,
\end{align}
\begin{align}\label{C}
C=\prod_{k=1}^{p}\left(A_{k}\right)^{A_{k}}\prod_{k=1}^{q}\left(B_{k}\right)^{-B_{k}},
\end{align}
\begin{align}\label{B}
B=(2\pi)^{\frac{q-p-1}{2}}C^{(1-\alpha)/m^{*}}\left(m^{*}\right)^{-1/2}\prod_{k=1}^{p}\left(A_{k}\right)^{-a_{k}+1/2}\prod_{k=1}^{m}\left(B_{k}\right)^{b_{k}-1/2}.
\end{align}

The series expansion of the Fox $H$-function~(\ref{H_integral}) is given by~\cite{MaHa08,MaSaHa10}
\begin{align}\label{H_expansion}
H_{p,q}^{m,n}\left[z\left|\begin{array}{l}
    (a_1,A_1),...,(a_p,A_p)\\
    (b_1,B_1),...,(b_q,B_q)
  \end{array}\right.\right]
=&\sum_{h=1}^{m}\sum_{k=0}^{\infty}\frac{\prod_{j=1, j\neq h}^{m}\Gamma\left(b_j-B_j\frac{b_h+k}{B_h}\right)\prod_{j=1}^{n}\Gamma\left(1-a_j+A_j\frac{b_h+k}{B_h}\right)}{\prod_{j=m+1}^{q}\Gamma\left(1-b_j+B_j\frac{b_h+k}{B_h}\right)\prod_{j=n+1}^{p}\Gamma\left(a_j-A_j\frac{b_h+k}{B_h}\right)}\nonumber\\&\times\frac{(-1)^kz^{(b_h+k)/B_h}}{k!B_h},
\end{align}
when the poles $\prod_{j=1}^{m}\Gamma\left(b_j-B_j\,s\right)$ are simple, i.e., $B_h(b_j+l)\neq B_j(b_h+k)$ for $j\neq h$, $h=1,\dots,m$, $l,k=0,1,2,\dots$.

%\section*{References}

\end{document}